# Relevance of Longitudinal Fields of Paraxial Optical Vortices

Kayn A. Forbes*, Dale Green, Garth A. Jones

*School of Chemistry, University of East Anglia, Norwich NR4 7TJ, United Kingdom*

Longitudinal electromagnetic fields generally become comparable with the usually dominant transverse components in strongly-focussed, non-paraxial beams. For optical vortex modes it is highlighted here how their angular momentum properties produce longitudinal fields that in general must be accounted for, even within the paraxial regime. First-order longitudinal components of quantized Laguerre-Gaussian modes are derived and numerically studied with respect to the paraxial parameter, highlighting light-matter and spin-orbit interactions that stem from longitudinal fields of weakly-focussed, paraxial beams in free space. New restrictions are cast on the validity of the paraxial approximation for optical vortices interacting with atoms, molecules and other nanostructures.

\* *k.forbes@uea.ac.uk*

*Introduction* – The idealized plane-wave solutions to the Maxwell and Helmholtz equations are more often than not utilized to provide a theoretical understanding of light-matter interactions [1]. Beyond the strictly transverse plane-wave solutions, the Helmholtz wave equation also permits 'laser beam' solutions, such as the ubiquitous Gaussian. Unlike the exact plane-wave solution, however, these are *approximate* beam-like solutions, and the ensuing solutions to the Maxwell equations are also only approximate. A key property of such solutions, however, is that they are not strictly transverse in free space, and the electric and magnetic fields generally have non-zero components parallel to the direction of propagation [2]: viz longitudinal fields.

In their seminal study, Lax *et al* [3] highlighted how the paraxial solutions to the scalar wave equation consist of a purely transverse zeroth-order field and smaller, first-order, longitudinal components whose magnitude for Gaussian-type beams depends on the paraxial factor $(kw_0)^{-1}$ where $k = 2\pi/\lambda$ is the wave number and $w_0$ the beam waist (at the focal point). Therefore, the importance of longitudinal fields for general laser modes are correlated to the degree of focussing, where strongly-focussed beams exhibit larger longitudinal components in their electromagnetic fields. For very strongly confined fields where $kw_0 \leq 1$, the legitimate separation of polarization and spatial degrees of freedom in scalar beams breaks down, and vector solutions to the wave equation that involve non-separable combinations of spatial and polarization modes are necessary [4,5]. The important role and application of longitudinal fields in vector beams has been well-established [6].

A realization that has led to a highly active area of modern optics is that longitudinal components to the total fields are not simply just quantitative corrections to the zeroth-order transverse fields but can exhibit highly distinct properties that influence propagation characteristics of the light or the ensuing interactions with matter. Focussed circularly-polarized Gaussian beams for example can induce orbital motion of trapped particles due to the spin-to-orbital angular momentum conversion [7,8].

Twisted light beams or optical vortices are an extremely well-studied type of structured laser light due to their rich angular momentum properties. For paraxial vortex modes, the individual photons can exhibit a spin angular momentum (SAM) $\sigma\hbar$, where $\sigma = \pm 1$, and orbital angular momentum (OAM) $\ell\hbar$, where $\ell \in \mathbb{Z}$, per photon. The widespread study and application of optical vortices is summarized in the following articles [9–11]. Most studies have been concerned with the angular momentum properties of non-paraxial and longitudinal fields of twisted light, such as spin-orbit interactions of light (SOI) and the transfer to particles to cause mechanical motion [12–14]. The application of twisted light in spectroscopic applications is a burgeoning area of research [11]. The potentially unique role that longitudinal fields can play in these applications has previously been highlighted for highly-focussed Laguerre-Gaussian (LG) beams [15,16]. A particularly important realization has been that the transfer of optical OAM to an atom [17] can only be correctly accounted for quantitatively if longitudinal fields of the input optical vortex are accounted for [18].

Here we derive the quantum electromagnetic field mode operators for LG beams that include first-order longitudinal field components in addition to the zeroth-order transverse fields. We highlight numerically that there are two distinct factors that influence the magnitude of longitudinal fields of optical vortices; firstly the well-known fact that a larger degree of focussing increases the longitudinal fields through a paraxial parameter $(kw_0)^{-1}$ weighting factor; but secondly that the angular momenta of optical vortices produces longitudinal fields that cannot be neglected for even weakly-focussed fields, highlighting both the quantitative and qualitative necessity of their inclusion even within the paraxial regime.

*Transverse and Longitudinal Fields of Laguerre-Gaussian modes* The most simple solutions to the wave equation are plane-waves, which subsequently become the transverse electromagnetic field solution to Maxwell's equations. In the Power-Zienau-Woolley (PZW) formulation of quantum electrodynamics (QED) [19,20] the electromagnetic field operators that couple to matter are the electric displacement field $d^\perp(r)$ and magnetic field $b(r)$. The superscript $\perp$ on $d^\perp(r)$ is with reference to the fact that for a neutral system $\nabla \cdot d = 0$ [21], and highlights that in the PZW formulation of QED all coulombic interactions are mediated by transverse photons [22]. The electric field displacement plane-wave mode expansion operator in the helicity basis is given as [22]

$$d^\perp_{x,y}(r) = \sum_{k,\sigma} \frac{\Omega}{\sqrt{2}} \left[ (\hat{x}+i\sigma\hat{y}) a^{(\sigma)}(k) e^{ik \cdot r} - H.c. \right], \quad (1)$$

where $k$ is the wave vector; $(\hat{x}+i\sigma\hat{y})$ is the polarization vector where $\sigma = \pm 1$ for left- and right-handed circularly-polarized light, respectively; $a^{(\sigma)}(k)$ is the annihilation operator; $\Omega$ is the normalization constant $\Omega = i(\hbar c k \varepsilon_0 / 2V)^{1/2}$ with $V$ the quantization volume; and $H.c.$ stands for the Hermitian conjugate. The subscript $x,y$ on $d^\perp_{x,y}(r)$ refers to the fact the fields are purely transverse *to the Poynting vector*, which for plane-waves in free space exactly coincides with the wave vector $k$.

Approximate beam-like solutions to the scalar Helmholtz equation which have finite transverse profiles are also possible. Solutions to the paraxial wave equation, which makes the approximation that the transverse profile changes very little in the propagation direction (usually defined as the $z$ axis) over a wavelength, form an ansatz for the Helmholtz equation when combined with a phase factor $e^{i(kz-\omega t)}$. Then electromagnetic beam-like solutions to Maxwell's equations can be found by ensuring all of Maxwell's laws are satisfied. Of course, one can take solutions to the full non-paraxial wave equation, one such case would be the Bessel beam class of optical vortices [23,24]. However, the most utilized of optical vortices are the Laguerre-Gaussian modes, which are solutions to the paraxial equation in cylindrical coordinates. As such, longitudinal components are of LG modes are generally neglected. The electric displacement field mode expansion operator for LG modes in the long Rayleigh range limit $z_R \gg z$ has previously been derived [25], and is given as

$$d^\perp_{x,y}(r) = \sum_{k,\sigma,\ell,p} \frac{\tilde{\Omega}}{\sqrt{2}} \left[ (\hat{x}+i\sigma\hat{y}) a^{(\sigma)}_{|\ell|,p}(k\hat{z}) f_{|\ell|,p}(r) e^{i(kz+\ell\phi)} - H.c. \right], \quad (2)$$

where $f_{|\ell|,p}(r) = \frac{C^{|\ell|}_p}{w_0} \left(\frac{\sqrt{2}r}{w_0}\right)^{|\ell|} e^{-\frac{r^2}{w_0^2}} L^{|\ell|}_p\left[\frac{2r^2}{w_0^2}\right]$ is a radial distribution function where $C^{|\ell|}_p$ is a constant and $L^{|\ell|}_p$ is the associated Laguerre polynomial, and $\tilde{\Omega}$ now includes the normalisation factor $A^{-3/2}_{\ell,p}$ for LG modes. In the terminology introduced by Lax et al. [3], (2) is a zeroth-order solution to Maxwell's equations, and thus completely transverse.

In the plane-wave solutions commonly used, the total field is simply $d^\perp(r) = d^\perp_{x,y}(r)$. However, for any beam-like solution to the Helmholtz equation, the zeroth-order (2) is only an approximation, and the total field is $d^\perp(r) = d^\perp_{x,y}(r) + \hat{z} d^\perp_z(r)$. The most direct method to calculate the first 'post-paraxial' longitudinal components is using the transversality conditions of Maxwell's equations [14,18]. In order to generate the first-order longitudinal terms for $d^\perp(r)$ we use Gauss's Law: $\nabla \cdot d = 0$. Thus the $z$ components $d^\perp_z(r)$ of the field can be determined via

$$\int \nabla^\perp \cdot d^\perp_{x,y}(r) dz = \sum_{k,\sigma,\ell,p} \frac{\tilde{\Omega}}{\sqrt{2}} \left[ \frac{i}{k}\left(\frac{\partial}{\partial x} + i\sigma\frac{\partial}{\partial y}\right) f_{|\ell|,p}(r) a^{(\sigma)}_{|\ell|,p}(k\hat{z}) e^{i(kz+\ell\phi)} - H.c. \right]. \quad (3)$$

Using Cartesian to cylindrical coordinate transformations produces the following mode expansion which now includes the additional first-order longitudinal corrections to the zeroth-order transverse field:

$$d^\perp(r) = \sum_{k,\sigma,\ell,p} \frac{\tilde{\Omega}}{\sqrt{2}} \left[ \left\{(\hat{x}+i\sigma\hat{y}) + \frac{i}{k}\left(\frac{\partial}{\partial r} - \ell\sigma\frac{1}{r}\right) e^{i\sigma\phi}\hat{z}\right\} \times f_{|\ell|,p}(r) a^{(\sigma)}_{|\ell|,p}(k\hat{z}) e^{i(kz+\ell\phi)} - H.c. \right]. \quad (4)$$

There are some interesting features of (4): For $\ell = 0$, one of the longitudinal terms is zero, but the other survives, highlighting how even longitudinal fields of a circularly-polarized Gaussian beam (i.e $LG^0_0$) exhibit a vortex of charge one structure in the $z$ direction through the phase factor $e^{i\sigma\phi}$. This SOI is known to occur in freely-propagating, *non-paraxial* beams of light [26]. Another SOI is evident when the incident beam does have an LG structure and accompanying OAM. For $\ell = \sigma$ the longitudinal fields form a vortex of charge 2; whereas for $\ell = -\sigma$ the beam exhibits a Gaussian structure with maximum intensity along the beam axis. This form of parallel and anti-parallel SAM and OAM projections, respectively, has been utilized in numerous studies [27,28].

For the vast majority of applications, the electric field is sufficient to describe light-matter interactions. However, magnetic interactions can become important in the correct settings, such as in chiral optics [29]. The magnetic field mode expansion is found using a similar approach, but with the aid of $\nabla \cdot b = 0$:

$$b(r) = \sum_{k,\sigma,\ell,p} \frac{\tilde{\Omega}}{\sqrt{2}c} \left[ \left\{(\hat{y}-i\sigma\hat{x}) + \frac{1}{k}\left(\sigma\frac{\partial}{\partial r} - \frac{\ell}{r}\right) e^{i\sigma\phi}\hat{z}\right\} \times f_{|\ell|,p}(r) a^{(\sigma)}_{|\ell|,p}(k\hat{z}) e^{i(kz+\ell\phi)} - H.c. \right]. \quad (5)$$

The first-order longitudinal components of the magnetic field (5) are wholly dependent on the angular momentum properties of the light, e.g. according to (5) a linearly-polarized Gaussian beam $(\sigma, \ell, p=0)$ possesses no longitudinal magnetic fields, only for strongly-focussed light where higher-order terms (second-order and above) contribute do longitudinal magnetic fields arise in this case. The longitudinal magnetic fields are $\pi/2$ out of phase with the longitudinal electric fields, a property previously recognized in being able to excite chiral nanostructues with linearly-polarized optical vortices [30,31].

*The Paraxial Approximation for Optical Vortices* The validity of the paraxial approximation has been questioned by numerous authors previously [4,32,33], and even with respect to optical vortices interacting with atoms [18,34]. These works predominantly concentrated on non-paraxial effects or specific systems. We now highlight how the transverse (zeroth-order) paraxial approximation is restrictive in general for optical vortices, with particular emphasis on weakly-focussed beams, and their interaction with nanostructures, such as atoms and molecules. To do this most explicilty and readily yield analytic results, it is best to look at LG modes where $p=0$, furthermore these are the most utilized modes in experiments.

The electric displacement field (4) takes on the following form when $p=0$ and the differentiation with respect to $r$ is carried out (remembering we are working within the long Rayleigh range limit; we have also dropped the obvious dependencies for notational clarity):

$$d^{\perp}(\mathbf{r}) = \sum_{k,\sigma,\ell} \frac{\tilde{\Omega}}{\sqrt{2}} \Big[ \big\{ (\hat{\mathbf{x}} + i\sigma\hat{\mathbf{y}}) + \frac{i}{k}\left(\frac{|\ell|}{r} - \frac{\ell\sigma}{r} - \frac{2r}{w_0^2}\right) e^{i\sigma\phi} \hat{\mathbf{z}} \big\} \\ \times f_{|\ell|,0} \, a \, e^{i(kz+\ell\phi)} - H.c. \Big]. \quad (6)$$

The magnitude of the last longitudinal term is weighted by the factor $2r/kw_0^2$, similarly to that of a Gaussian beam, where $kw_0^2$ is sometimes referred to as the diffraction length [3]. This generally small contribution (for weakly-focussed beams [35]) to the longitudinals fields is independent of the angular momentum properties of the field, and is exhibited by any field mode which possesses a Gaussian factor.

However, there are an additional two terms that depend on the angular momentum properties of the beam. First there is the contribution dependent on the factor $|\ell|/kr$ which is the absolute value of the skew angle of the Poynting vector at a given location [36]. If the light also posesses SAM, there is the additional skew-angle-like term dependent on $\ell\sigma/kr$, which thus includes the mixing of helicity and topological charge values. The skew angle term dependent on $|\ell|$ reveals that a longitudinal field exists for a non-circularly polarized ($\sigma=0$) LG beam, whilst the second term signififes a SOI of light in freely-propagating circularly polarized optical vortices in free-space, a phenomenon highlighted some time ago [37] but has seemingly received relatively little attention.

The notion that longitudinal fields for optical vortices may be neglected unless the light is non-paraxial pervades the literature, e.g. [14,38–40], and is a commonly held misconception, likely stemming from the fact that this is generally true for laser modes that do not possess OAM. The intensity distributions as a function of beam waist of a variety of LG modes where the first-order longitudinal fields have been included with the generally used zeroth-order tranvserse fields is given in Figure 1 and Figure 2.

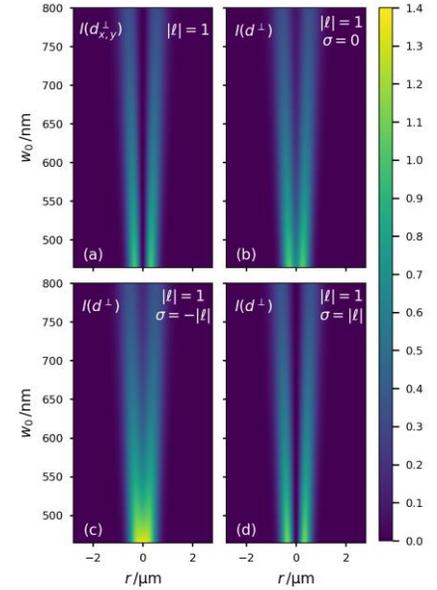

**Figure 1**: Intensity distributions for an LG beam with $|\ell|=1$ for a changing beam waist (wavelength 729nm) a) the intensity contains only the transverse field components b) the total intensity including longitudinal contributions for $\sigma=0$ c) Total intensity for antiparallel OAM and SAM and d) Total intensity for parallel OAM and SAM. The range of beam waists used in Figure 1 and 2 correspond to $4 \leq kw_0 \leq 20$.

The range of beam waists used in Figure 1 and 2 correspond to $4 \leq kw_0 \leq 20$, i.e. *within* the paraxial range [32]. Just as it is well-established that longitudinal fields cannot be neglected when unstructured (non-OAM) light is strongly focussed because terms dependent on factors like $2r/kw_0^2$ become important, Figure 1 shows that certain contributions to longitudinal fields for optical vortices cannot likewise be neglected, but even for beams that would be considered to be propagating paraxially.

The importance of these fields are still bound to specific scenarios due to the dependence on the radial distribution function, beam waist, optical angular momentum, and in comparison to the zeroth-order transverse fields are weighted by the wavelength (inverse wave number).

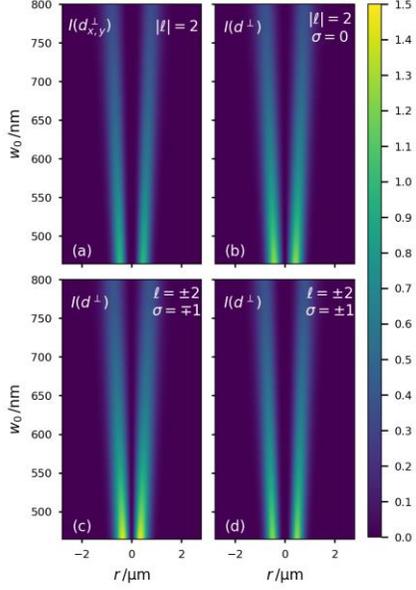

**Figure 2**: Intensity distributions for an LG beam with $|\ell|=2$ for a changing beam waist (wavelength 729nm) a)-d) correspond to the same conditions as those in Figure 1, the range of $kw_0$ is also the same.

Whilst it is true that these terms do indeed become larger the more focussed the beam is, Figure 1b and 1c clearly highlights how they manifest even in LG modes that are weakly focussed, particularly in the so-called 'vortex core' which evidently is not truly empty for a range of parameters which fall within the paraxial regime. Furthermore, if the twisted light is also circularly-polarized the effects become more significant than if $\sigma=0$, and in the case of anti-parallel SAM and OAM the paraxial approximation significantly fails for the whole range of values of $kw_0$ (Figure 1c). As Figure 2 shows, for higher values of $|\ell|$ the on-axis intensity distribution vanishes for paraxial beams, and the total field intensity distribution much more resembles the transverse only components, but the transverse only description still differs from the total field, especially for the lower values of $kw_0$ when $\ell=\pm2, \sigma=\mp1$ as in Figure 2c. Although we have shown that the inclusion of solely zeroth-order transverse fields is not in general adequate for paraxial LG modes, it is clear that the paraxial approximation most significantly fails quantitatively for $\ell=\pm1, \sigma=0$ and $\ell=\pm1, \sigma=\mp1$.

*Application: Single-photon absorption.* The previous section shows us longitudinal fields are important to account for in spectroscopies as the highly-position-dependent intensity structure will need to be properly accounted for.

Single-photon absorption is the simplest optical process, and is therefore an appropriate initial case to investigate the role of longitudinal fields in light-matter interactions. We may calculate the matrix element (or quantum amplitude) $M_{FI}$ of single-photon absorption in the electric dipole approximation using standard time-dependent perturbation methods, with $M_{FI}$ given by: $\langle F|H_{int}|I\rangle = \langle F|-\varepsilon_0^{-1}\boldsymbol{\mu}\cdot\boldsymbol{d}^\perp(\boldsymbol{r})|I\rangle$, where the initial state vector $|I\rangle = |n(k,\sigma,\ell,p); E_0\rangle$ consists of $n$ photons in the LG mode $(k,\sigma,\ell,p)$ and the material in the ground state $E_0$; the final state is given by $|F\rangle = |(n-1)(k,\sigma,\ell,p); E_\alpha\rangle$, where the mode has lost a photon and the material now exists in the excited state denoted by $\alpha$; $\boldsymbol{\mu}$ is the electric-dipole transition moment operator.

Using the mode expansion for the electric displacement field operator (4), the matrix element is

$$M_{FI} = -\tilde{\Omega}\sqrt{n}\left[e_i^{L/R} + \frac{i}{\sqrt{2}k}\left(\frac{\partial}{\partial r} - \frac{\ell\sigma}{r}\right)\hat{z}_i e^{i\sigma\phi}\right] f \mu_i^{\alpha 0} e^{i(kz+\ell\phi)}, \quad (7)$$

where we now use suffix notation for tensor quantities and $e_i^{L/R} = 2^{-1/2}(\hat{x}+i\sigma\hat{y})_i$ is the polarization vector for circularly polarized light. As is standard [22], the matrix element is inserted into the Fermi rule to yield the rate $\Gamma$ of single-photon absorption as:

$$\Gamma \propto |M_{FI}|^2 = n\tilde{\Omega}^2 \left[ e_i^{L/R}\bar{e}_j^{L/R} f^2 \right.$$
$$+ \frac{1}{kr}(\sigma r f f' - \ell f^2)\hat{z}_i(\hat{y}_j\cos\phi - \hat{x}_j\sin\phi) \quad (8)$$
$$\left. + \frac{\hat{z}_i\hat{z}_j}{2k^2r^2}\left(-2\ell\sigma r f f' + (rf')^2 + (\ell f)^2\right)\right]\mu_i^{\alpha 0}\bar{\mu}_j^{\alpha 0}.$$

The prime superscripts in (8) denote partial differentiation with respect to $r$. The first term in square brackets in (8) is the standard rate of absorption via the purely transverse zeroth-order electric field, and it is therefore evident through the multitude of additional terms that accounting for longitudinal fields offer numerous additional optical interactions and qualitative corrections to the zeroth-order fields. The terms dependent on $\phi$ stem from the interferences between the transverse and longitudinal fields, and if a full beam-profile integration over $\phi$ is carried out (i.e. $\phi_0 \to \phi_0 + 2\pi$) these effects vanish. To render these observable the signals stemming from individual nanostructures or sub-domains must be resolved. The final terms are all purely longitudinal in nature, and their importance is determined by the factors discussed in the previous section.

The rate of absorption given by (8) specifically corresponds to a nanostructure with a fixed orientation with respect to the input optical axis. It is clear that that the zeroth-order transverse fields can excite electric dipole transitions which must have allowed components in $x,y$-directions, whereas excitation through the pure longitudinal terms excite transitions which must exhibit components along the direction of propagation $z$. A similar scheme was used to map the fluorescence of molecules with specific orientations in order to precisely determine of the structure of electromagnetic fields for vector vortex beams [41]. It is also worthwhile noting that individual contributions to the total rate depend on $\sigma$, i.e. the handedness of the input circular polarization; the sign of $\ell$, i.e. the handedness of the optical vortex; and the product of the two $\ell\sigma$. Generally materials, such as molecular matter, need to be chiral in order to exhibit differential effects with respect to the sign of $\sigma$ (optical activity) through higher-order magnetic dipole and electric quadrupole interferences with the electric dipole transitions, and are thus usually weak effects [42]. The rate (8) tells us that in principle it is possible that photon absorption through purely electric-dipole transitions can yield a small differential rate for optical vortices with different helical wavefront and polarization handedness. These differential rates that depend on the optical handedness here are comparable to standard optical activity of circularly polarized light, but in comparison to these traditional chiroptical interactions that probe the local helicity of light, the phenomena here are clearly spatial effects related to a radially varying and optical angular momentum-dependent intensity structure [43].

Orientational averaging is done using standard methods [44], namely for a second rank tensor $\langle \mu_i \bar{\mu}_j \rangle = \frac{1}{3}\delta_{ij}\boldsymbol{\mu}^{\alpha 0} \cdot \bar{\boldsymbol{\mu}}^{\alpha 0}$:

$$\langle \Gamma \rangle \propto \frac{n\tilde{\Omega}^2}{3}\left( f^2 - \frac{\ell\sigma}{k^2 r} ff' + \frac{1}{2k^2} f'^2 + \frac{\ell^2}{2k^2 r^2} f^2 \right)|\boldsymbol{\mu}^{\alpha 0}|^2.$$

(9)

The total *averaged* rate (9) and its individual components are plotted in Figures 3 and 4. It is interesting to note how an orientational average of the individual material particles also leads to the interference terms averaging to zero. Clearly the different material transition moment orientational dependences of (8) to the transverse and longitudinal fields are also lost. The SOI term dependent on $\ell\sigma$ still survives, so that even randomly oriented molecules will still exhibit a small difference in the rate of absorption depending on the polarization and wavefront handedness – this is most clear by comparing Figure 3b and c for a given position $r$ (though note how the light must possess both SAM and OAM in this case).

It is clear to see that the rate of absorption has an acute dependence on radial position, but also the total rate is altered by the longitudinal fields even for beams propagating within the paraxial regime. The results displayed in Figure 3 corresponds to a laser beam propagating at the limit of the paraxial approximation $kw_0 \approx 4$: Figure 4 highlights the same rate but for a beam well within the paraxial regime with $kw_0 \approx 7$. Note that Quinteiro *et al.* [18] highlighted how longitudinal fields in the specific case of parallel and anti-parallel SAM and OAM had a significant influence on atomic electric quadrupole transitions even when $kw_0 \approx 23$ (see our Figure 1c).

Importantly, for a paraxial beam of any structure, it is to be expected that the longitudinal fields would be smaller than the transverse fields (unlike the scenarios that can occur in highly-focussed beams where the longitudinal fields can dominate), however it is clear that they cannot be neglected for certain situations with paraxial vortices.

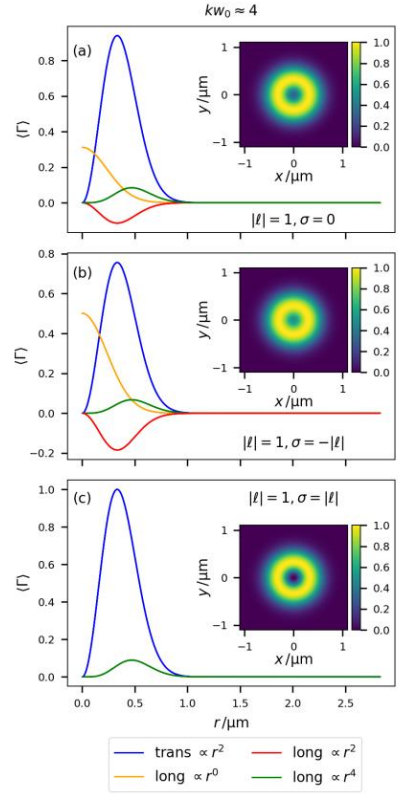

**Figure 3**: 1D Plots of the individual contributions to the rate of single-photon absorption (Eq.(9)) with 2D inset plot corresponding to total rate ($p = 0$, wavelength 729nm, beam waist 470nm: $kw_0 \approx 4$.

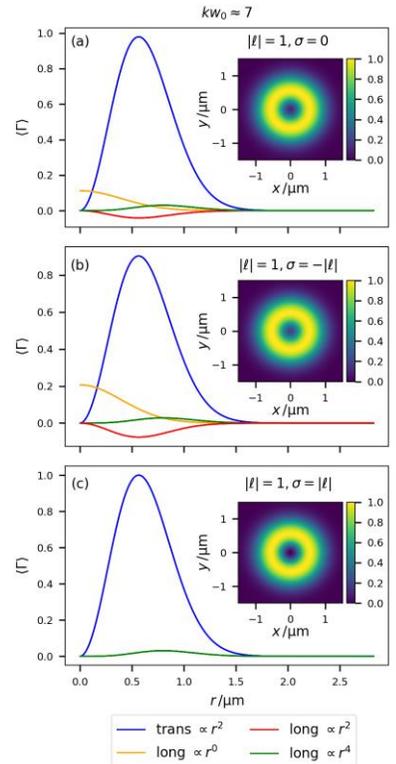

**Figure 4**: 1D Plots of the individual contributions to the rate of single-photon absorption (Eq.(9)) with 2D inset plot corresponding to total rate ( $p=0$, wavelength 729nm, beam waist 800nm: $kw_0 \approx 7$.

*Discussion and Conclusion* LG modes are solutions to the paraxial wave equation, and therefore must strictly be bound to any approximations associated with paraxially propagating light. Detailed conditions of where the paraxial approximation fails for general laser modes can be found in [32,33]. If one wishes to rigorously account for the longitudinal fields of *strongly-focussed* LG beams, then either (a) the non-paraxial LG solutions should be utilised [23,24], (b) a systematic expansion in paraxial parameter in the spirit of previous studies [3,45,46] or (c) explicilt inclusion of the focussing approach with high NA [13,27,47,48] should be adopted. Importantly, as we have shown, the other terms dependent on the angular momentum properties can be important even for weakly-focussed optical vortices, and their first-order nature means their contributions are completely valid within the paraxial regime.

It is worth discussing a subtle point about the magnetic field (5) which was derived using the divergenceless nature of magnetic fields. The magnetic field could have been derived using Faraday's law $\nabla \times e^{\perp}(r) = -\partial b(r)/\partial t$ which yields the fields as in (5) plus higher-order corrections to the transverse field which essentially stem from $\partial^2/\partial z^2$ variations of the amplitude. Working within the paraxial approximation these terms have been neglected here due to the fact they only become important in situations where a paraxial description of the field would be innacurate. It has been highlighted how for highly-focussed vortex beams they exhibit unique features for $|\ell|=2$ [16]. Similarly, it is worth making the point that restricting our results to within the Rayleigh range is legitimate for LG modes as the *z* dependence of the full, non-Rayleigh range LG mode function only leads to longitudinal and transverse fields which are higher-order in the paraxial parameter (i.e. second-, third-, and so on post-paraxial approximations), and thus are actually beyond the validity of the paraxial wave equation and so their inclusion is not supported here either.

The field of twisted light and optical OAM has largely been concerned with applications in mechanical nanomanipulation, communications, and quantum information studies – only in the last few years have the unique properties of twisted beams been implemented in atomic and molecular optics and spectroscopy. Here we have highlighted how in such studies, longitudinal fields of optical vortices must be accounted for even if the beam is considered to be propagating paraxially. Indeed, any application of weakly-focussed vortices clearly require the inclusion of longitudinal fields, particularly for low values of $\ell$ and nanostructures placed close to the so-called vortex singularity. The cases which show most deviation from the purely zeroth-order transverse field description of LG modes appear to be $|\ell|=1, \sigma=0$; $\ell=\pm 1, \sigma=\mp 1$, $\ell=\pm 2, \sigma=\mp 1$, particularly the first two. Nanostructures with specific orientation with respect to the optical axis have more potential to exhibit larger and more interesting effects with the longitudinal fields than systems of randomly oriented structures. Unlike the strongly-focussed case of non-paraxial beams where the longitudinal components can dominate the transverse fields, here it has been shown that for paraxial optical vortices the inclusion of first-order longitudinal fields is still important both qualitatively and quantitatively, as they introduce novel optical interactions with matter as well as alter the electromagnetic fields and corresponding position-dependent intensity structure.

K.A.F. thanks the Leverhulme Trust for funding through a Leverhulme Early Career Fellowship ECF-2019-398.

[1] G. Grynberg, A. Aspect, and C. Fabre, *Introduction to Quantum Optics: From the Semi-Classical Approach to Quantized Light* (Cambridge University Press, Cambridge, U.K., 2010).
[2] A. Zangwill, *Modern Electrodynamics* (Cambridge University Press, Cambridge, 2013).
[3] M. Lax, W. H. Louisell, and W. B. McKnight, *From Maxwell to Paraxial Wave Optics*, Phys. Rev. A **11**, 1365 (1975).
[4] C. G. Chen, P. T. Konkola, J. Ferrera, R. K. Heilmann, and M. L. Schattenburg, *Analyses of Vector Gaussian Beam Propagation and the Validity of Paraxial and Spherical Approximations*, JOSA A **19**, 404 (2002).
[5] L. Novotny and B. Hecht, *Principles of Nano-Optics* (Cambridge University Press, Cambridge, 2012).
[6] Q. Zhan, *Cylindrical Vector Beams: From Mathematical Concepts to Applications*, Adv. Opt. Photonics **1**, 1 (2009).
[7] H. Adachi, S. Akahoshi, and K. Miyakawa, *Orbital Motion of Spherical Microparticles Trapped in Diffraction Patterns of Circularly Polarized Light*, Phys. Rev. A **75**, 063409 (2007).
[8] Y. Zhao, D. Shapiro, D. McGloin, D. T. Chiu, and S. Marchesini, *Direct Observation of the Transfer of Orbital Angular Momentum to Metal Particles from a Focused Circularly Polarized Gaussian Beam*, Opt. Express **17**, 23316 (2009).
[9] H. Rubinsztein-Dunlop et al., *Roadmap on Structured Light*, J. Opt. **19**, 013001 (2016).
[10] S. M. Barnett, M. Babiker, and M. J. Padgett, editors, *Optical Orbital Angular Momentum* (The Royal Society, London, 2017).
[11] M. Babiker, D. L. Andrews, and V. E. Lembessis, *Atoms in Complex Twisted Light*, J. Opt. **21**, 013001 (2018).
[12] T. A. Nieminen, A. B. Stilgoe, N. R. Heckenberg, and H. Rubinsztein-Dunlop, *Angular Momentum of a Strongly Focused Gaussian Beam*, J. Opt. Pure Appl. Opt. **10**, 115005 (2008).
[13] P. B. Monteiro, P. A. M. Neto, and H. M. Nussenzveig, *Angular Momentum of Focused Beams: Beyond the Paraxial Approximation*, Phys. Rev. A **79**, 033830 (2009).
[14] K. Y. Bliokh and F. Nori, *Transverse and Longitudinal Angular Momenta of Light*, Phys. Rep. **592**, 1 (2015).

[15] V. Klimov, D. Bloch, M. Ducloy, and J. R. R. Leite, *Detecting Photons in the Dark Region of Laguerre-Gauss Beams*, Opt. Express **17**, 9718 (2009).

[16] V. V. Klimov, D. Bloch, M. Ducloy, and J. R. Leite, *Mapping of Focused Laguerre-Gauss Beams: The Interplay between Spin and Orbital Angular Momentum and Its Dependence on Detector Characteristics*, Phys. Rev. A **85**, 053834 (2012).

[17] C. T. Schmiegelow, J. Schulz, H. Kaufmann, T. Ruster, U. G. Poschinger, and F. Schmidt-Kaler, *Transfer of Optical Orbital Angular Momentum to a Bound Electron*, Nat. Commun. **7**, 12998 (2016).

[18] G. F. Quinteiro, F. Schmidt-Kaler, and C. T. Schmiegelow, *Twisted-Light–Ion Interaction: The Role of Longitudinal Fields*, Phys. Rev. Lett. **119**, 253203 (2017).

[19] D. L. Andrews, G. A. Jones, A. Salam, and R. G. Woolley, *Perspective: Quantum Hamiltonians for Optical Interactions*, J. Chem. Phys. **148**, 040901 (2018).

[20] D. L. Andrews, D. S. Bradshaw, K. A. Forbes, and A. Salam, *Quantum Electrodynamics in Modern Optics and Photonics: Tutorial*, JOSA B **37**, 1153 (2020).

[21] M. Babiker, E. A. Power, and T. Thirunamachandran, *On a Generalization of the Power–Zienau–Woolley Transformation in Quantum Electrodynamics and Atomic Field Equations*, Proc. R. Soc. Lond. Math. Phys. Sci. **338**, 235 (1974).

[22] D. P. Craig and T. Thirunamachandran, *Molecular Quantum Electrodynamics: An Introduction to Radiation-Molecule Interactions* (Courier Corporation, New York, 1998).

[23] S. M. Barnett and L. Allen, *Orbital Angular Momentum and Nonparaxial Light Beams*, Opt. Commun. **110**, 670 (1994).

[24] D. L. Andrews and M. Babiker, editors, *The Angular Momentum of Light* (Cambridge University Press, Cambridge, 2012).

[25] L. D. Romero, D. L. Andrews, and M. Babiker, *A Quantum Electrodynamics Framework for the Nonlinear Optics of Twisted Beams*, J. Opt. B Quantum Semiclassical Opt. **4**, S66 (2002).

[26] K. Y. Bliokh, F. J. Rodríguez-Fortuño, F. Nori, and A. V. Zayats, *Spin–Orbit Interactions of Light*, Nat. Photonics **9**, 796 (2015).

[27] Y. Iketaki, T. Watanabe, N. Bokor, and M. Fujii, *Investigation of the Center Intensity of First-and Second-Order Laguerre-Gaussian Beams with Linear and Circular Polarization*, Opt. Lett. **32**, 2357 (2007).

[28] Y. Zhao, J. S. Edgar, G. D. Jeffries, D. McGloin, and D. T. Chiu, *Spin-to-Orbital Angular Momentum Conversion in a Strongly Focused Optical Beam*, Phys. Rev. Lett. **99**, 073901 (2007).

[29] D. L. Andrews, *Quantum Formulation for Nanoscale Optical and Material Chirality: Symmetry Issues, Space and Time Parity, and Observables*, J. Opt. **20**, 033003 (2018).

[30] C. Rosales-Guzmán, K. Volke-Sepulveda, and J. P. Torres, *Light with Enhanced Optical Chirality*, Opt. Lett. **37**, 3486 (2012).

[31] P. Woźniak, I. D. Leon, K. Höflich, G. Leuchs, and P. Banzer, *Interaction of Light Carrying Orbital Angular Momentum with a Chiral Dipolar Scatterer*, Optica **6**, 961 (2019).

[32] S. Nemoto, *Nonparaxial Gaussian Beams*, Appl. Opt. **29**, 1940 (1990).

[33] P. Vaveliuk, B. Ruiz, and A. Lencina, *Limits of the Paraxial Approximation in Laser Beams*, Opt. Lett. **32**, 927 (2007).

[34] G. F. Quinteiro, C. T. Schmiegelow, and F. Schmidt-Kaler, *The Paraxial Approximation Fails to Describe the Interaction of Atoms with General Vortex Light Fields*, ArXiv Prepr. ArXiv200400040 (2020).

[35] A. Carnicer, I. Juvells, D. Maluenda, R. Martínez-Herrero, and P. M. Mejías, *On the Longitudinal Component of Paraxial Fields*, Eur. J. Phys. **33**, 1235 (2012).

[36] J. Leach, S. Keen, M. J. Padgett, C. Saunter, and G. D. Love, *Direct Measurement of the Skew Angle of the Poynting Vector in a Helically Phased Beam*, Opt. Express **14**, 11919 (2006).

[37] L. Allen, V. E. Lembessis, and M. Babiker, *Spin-Orbit Coupling in Free-Space Laguerre-Gaussian Light Beams*, Phys. Rev. A **53**, R2937 (1996).

[38] E. Santamato, *Photon Orbital Angular Momentum: Problems and Perspectives*, Fortschritte Phys. Prog. Phys. **52**, 1141 (2004).

[39] J. Wang, F. Castellucci, and S. Franke-Arnold, *Vectorial Light–Matter Interaction: Exploring Spatially Structured Complex Light Fields*, AVS Quantum Sci. **2**, 031702 (2020).

[40] O. V. Angelsky, A. Y. Bekshaev, S. G. Hanson, C. Y. Zenkova, I. I. Mokhun, and Z. Jun, *Structured Light: Ideas and Concepts*, Front. Phys. **8**, (2020).

[41] L. Novotny, M. R. Beversluis, K. S. Youngworth, and T. G. Brown, *Longitudinal Field Modes Probed by Single Molecules*, Phys. Rev. Lett. **86**, 5251 (2001).

[42] L. D. Barron, *Molecular Light Scattering and Optical Activity* (Cambridge University Press, Cambridge, 2009).

[43] K. A. Forbes and D. L. Andrews, *Orbital Angular Momentum of Twisted Light: Chirality and Optical Activity*, J. Phys. Photonics (2021).

[44] D. L. Andrews and T. Thirunamachandran, *On Three-Dimensional Rotational Averages*, J. Chem. Phys. **67**, 5026 (1977).

[45] G. P. Agrawal and D. N. Pattanayak, *Gaussian Beam Propagation beyond the Paraxial Approximation*, JOSA **69**, 575 (1979).

[46] M. Couture and P.-A. Belanger, *From Gaussian Beam to Complex-Source-Point Spherical Wave*, Phys. Rev. A **24**, 355 (1981).

[47] N. Bokor, Y. Iketaki, T. Watanabe, and M. Fujii, *Investigation of Polarization Effects for High-Numerical-Aperture First-Order Laguerre-Gaussian Beams by 2D Scanning with a Single Fluorescent Microbead*, Opt. Express **13**, 10440 (2005).

[48] K. Y. Bliokh, E. A. Ostrovskaya, M. A. Alonso, O. G. Rodríguez-Herrera, D. Lara, and C. Dainty, *Spin-to-Orbital Angular Momentum Conversion in Focusing, Scattering, and Imaging Systems*, Opt. Express **19**, 26132 (2011).